\documentclass{PoS}
\usepackage{amsmath}
\usepackage{amssymb}
\usepackage[Gray,squaren,thinqspace,thinspace]{SIunits}

\newcommand{\lms}{\Lambda_{\overline{\mbox{\tiny{MS}}}}}

\title{The interplay between the $\langle A_\mu^2\rangle$ condensate and instantons}

\ShortTitle{The interplay between the $\langle A_\mu^2\rangle$ condensate and instantons}

\author{\speaker{David Vercauteren}\\
        Universitat de Val\`encia\\
        E-mail: \email{david.vercauteren@uv.es}}

\author{Henri Verschelde\\
        Ghent University\\
        E-mail: \email{henri.verschelde@ugent.be}}

\abstract{Using the Local Composite Operator formalism, we analytically study the dimension two gluon condensate in the presence of instantons. We first use the dilute gas approximation and partially solve the infrared problem of instanton physics. In order to find quantitative results, however, we turn to an instanton liquid model, where we find a two-component picture of the condensate: one component comes from instantons, a second component is non-perturbatively generated by quantum fluctuations around the instantons.}

\FullConference{The many faces of QCD\\
		November 1-5, 2010\\
		Gent Belgium}

\begin{document}

\section{Introduction}
\subsection{The dimension two condensate}
Past years have witnessed a great deal of interest in the possible existence of mass dimension two condensates in gauge theories as proposed in \cite{Gubarev:2000eu,Gubarev:2000nz}, see for example \cite{Boucaud:2001st, RuizArriola:2006gq, Verschelde:2001ia, Vercauteren:2007gx, Vercauteren:2010rk, Dudal:2005na, Dudal:2003vv, Browne:2003uv, Furui:2005he, Gubarev:2005it, Chernodub:2008kf, Andreev:2006vy} and references therein for approaches based on phenomenology, operator product expansion, lattice simulations, an effective potential and the string perspective.

The obvious question of the gauge invariance of $\langle A_\mu^2\rangle$ can be resolved by considering its minimum over the gauge orbit
\begin{equation} \label{mingaugeorbit}
A_{\min}^2 = \min_{U\in SU(N)} V^{-1} \int d^4x (A_\mu^U)^2 \;,
\end{equation}
which is gauge invariant by construction, albeit very non-local. It should be mentioned
that obtaining the global minimum is delicate
due to the problem of Gribov ambiguities
\cite{Gribov:1977wm}. As is well-known, local gauge
invariant dimension two operators do not exist in
Yang--Mills gauge theories. The non-locality of \eqref{mingaugeorbit}
is best seen when it is expressed as \cite{Lavelle:1995ty}\footnote{We will always work in Euclidean space-time.}
\begin{equation} \label{akwadraatlang}
A_{\min}^{2} = \int d^{4}x\left[ A_{\mu }^{a}\left( \delta _{\mu \nu }-\frac{\partial _{\mu }\partial _{\nu }}{\partial ^{2}}\right) A_{\nu}^{a} - gf^{abc}\left( \frac{\partial _{\nu}}{\partial ^{2}}\partial A^{a}\right) \left( \frac{1}{\partial^{2}}\partial {A}^{b}\right) A_{\nu }^{c}\right] + \mathcal O(A^{4}) \;.
\end{equation}
All efforts so far have concentrated on the Landau
gauge $\partial_\mu A_\mu = 0$. The preference for this particular
gauge fixing is obvious since the non-local expression
\eqref{akwadraatlang} reduces to a local operator, more precisely
\begin{equation}
\partial_\mu A_\mu = 0 \Rightarrow A^2_\text{min} = A_\mu^2 \;.
\end{equation}

The relevance of the condensate $\langle A_\mu^2\rangle_\mathrm{min}$ was discussed in \cite{Gubarev:2000eu}, where it was shown that it can serve as a measure for monopole condensation, and hence for confinement, in the case of compact QED. In QCD, however, the situation is slightly more complicated and in \cite{Gubarev:2000nz} a two-component picture was proposed. There is a soft (infrared) part, which contributes $1/q^2$ power corrections to the Operator Product Expansion of (gauge variant) quantities such as the gluon propagator (see e.g. \cite{Boucaud:2001st, Lavelle:1988eg}), and a hard part which can enter physical correlator and is modeled in phenomenology by gluon masses (see e.g. \cite{RuizArriola:2006gq, Megias:2005ve}). This last part can be seen as an example of non-perturbative ultraviolet effects in gauge theory. These gluon masses are nothing new, see for example \cite{Cornwall:1981zr, Parisi:1980jy}. A gluon mass will lead to violation of unitarity, but one can hope that this will be solved when, somehow, confinement is correctly taken into account.

\subsection{Instantons}
Instantons play an important role in the QCD vacuum and have a large influence in many infrared properties (see \cite{Schafer:1996wv} for a review). As such it is an interesting question what their connection with the dimension two condensate is. From the analytic viewpoint, however, instantons are not easily handled, as, for example, 't~Hooft's seminal calculation in the dilute gas approximation found an infrared instability \cite{thooft}. Later models include the liquid models, see e.g. \cite{Shuryak:1981ff}.

A first study concerning the connection between instantons and the $\langle A_\mu^2\rangle$ condensate has been done on the lattice by Boucaud \emph{et al.} \cite{Boucaud:2002nc,Boucaud:2002wy}, and a rather large instanton contribution to the condensate has been found, showing some agreement with the results from an OPE approach to the gluon propagator from \cite{Boucaud:2001st}. However, the condensate may get separate contributions from other sources, as for example the non-perturbative high-energy fluctuations leading to the condensate found in \cite{Verschelde:2001ia}. The opposite viewpoint is just as interesting: what is the influence of an effective gluon mass on the instanton ensemble? In 't~Hooft's seminal paper he found that, in a Higgs model, a gauge boson mass stabilizes the instanton gas.

\subsection{The LCO formalism}
In order to study the condensation of a Local Composite Operator (LCO) analytically, we turn to the formalism developed by Verschelde \emph{et al.} \cite{Verschelde:2001ia}. One starts by adding a source $J$ coupled to $A_\mu^2$ to the Lagrangian density. Doing so, however, makes the theory non-renormalizable at the quantum level. To solve this, a term quadratic in the source must be added, which in turn spoils the energy interpretation of the effective action. One way around this is to perform the Legendre inversion, but this is rather cumbersome, especially with a general, space-time-dependent source. One can also use a Hubbard--Stratonovich transform, which introduces an auxiliary field (whose interpretation is just the condensate) and eliminates the term quadratic in the source. Also, the coefficient of the term quadratic in the source is a new, arbitrary parameter, which has to be chosen somehow. By demanding it to be a meromorphic function of the gauge coupling, the renormalization group gives this parameter a unique value. Details can be found in \cite{Verschelde:2001ia}. The result was that the Yang--Mills vacuum favors a finite value for the expectation value $\langle A_\mu^2\rangle$\footnote{This value is negative, contrarily to what one might naively expect. As the condensate is actually divergent, an infinte quantity is subtracted during the renormalization, leaving a finite but negative value.}. The precise renormalization details of the procedure were given in \cite{Dudal:2002pq}.

An extra question arising is which gauge to choose. All instanton calculations are done in background gauges, as analytic computations in non-background gauges are quite impossible. The LCO formalism does not give classical fields a mass in the Landau background gauge, however. In the electroweak theory considered by 't~Hooft in \cite{thooft} it is exactly this classical mass which suppresses large instantons by the simple fact that large instantons are no solutions to the massive field equations anymore, while small instantons can still be considered approximate solutions. In the following sections both gauges will be studied.

\section{Ordinary Landau gauge}
\subsection{Discussion}
If we want to have a mass already at the classical level, it is necessary to work in the non-background Landau gauge. Although the computations cannot be carried through in this gauge, it still possible to find the qualitative form of the result. In order to circumvent the question of which background to take for the $\sigma$ field\footnote{Allowing $\sigma$ to obey its own \emph{classical} field equations does not lead to non-trivial results.}, it is more opportune to start before the point where the Hubbard--Stratonovich transformation is introduced.

We start from
\begin{equation}
-\frac12\langle A_\mu^2\rangle = \left. \frac\delta{\delta J} \ln \int [dA_\mu] e^{-S-\frac12JA_\mu^2+\frac\zeta2J^2} \right|_{J=0} \;.
\end{equation}
As the source is small, instantons will be approximate solutions. Eventually, we can correct the instanton using the valley method \cite{Affleck:1980mp}, but this turns out not to give more insight. From renormalization group arguments, we can now write down the general form of the one-loop result:
\begin{equation}
W[J] = W^{0I}[J] - \int_0^\infty \frac{d\rho}{\rho^5} \exp\left(-\frac{8\pi^2}{g^2} - \frac{6\pi^2}{g^2}J\rho^2 + \frac{11}3\ln(\mu^2\rho^2) + f_1(J\rho^2) + \cdots\right) \;,
\end{equation}
where the dilute instanton gas approximation has been used, giving an exponential of the instanton contribution. Here, $W^{0I}$ denotes the zero-instanton result, and $f_1$ is an unknown function which contains the quantum corrections. A factor of the space-time volume has been left out. For finite $J$, the integral over the instanton size $\rho$ is now convergent and can be done, and after the Legendre inversion one finds
\begin{equation}
\Gamma[\sigma] = \Gamma^{0I}[\sigma] - g^{10/3}\mu^{22/3}\sigma^{-5/3} e^{-\frac{8\pi^2}{g^2}} f_2(g^2) \;,
\end{equation}
where $f_2$ is yet another unknown function, and $\Gamma^{0I}$ is the zero-instanton result. If the coupling is sufficiently small, the instanton correction can be ignored and the zero-instanton result is recovered. The instanton term can then be considered as a small perturbation, slightly shifting the value of the condensate. However, no matter how small the coupling, the second term will always diverge for sufficiently small $\sigma$, and so the effective action will be unbounded from below\footnote{It is easy to see that $f_2(g^2)$ must be positive, at least for small $g^2$.}. This is of course related to the infrared divergence found in the case without condensate.

\subsection{Conclusion}
Two problems can be identified. First there is the resilience of the infrared divergence. One could say this is due to the strength of the LCO formalism ---the gluon mass is left free in order to determine it by its gap equation, which allows the possibility for the mass to be zero, which again allows instantons to proliferate and to so destabilize the action. This can be solved invoking only a little hand-waving: when $\sigma$ is small, the dilute instanton gas approximation is not valid, and so this part of the result must be thrown away. The final conclusion is that instantons slightly shift the value of $\langle A_\mu^2\rangle$.

This leaves a second problem: one would expect each instanton to give a contribution of $12\pi^2\rho^2/g^2$ to the condensate already at the classical level. This does not happen, which is due to the way the problem has been approached. The dilute instanton gas approximation starts from the one-instanton contribution and exponentiates it to give a gas. The contribution of one instanton to the condensate is negligible ---it is finite, while the total condensate is proportional to the space-time volume--- and so it drops out.

In the background gauge this last problem is readily solved: the classical and quantum mechanical contributions are neatly separated from the start. Furthermore it turns out that the computations can all be done, which allows for a quantitative result to be given as well. Only the infrared divergence still remains as a problem, but, as some hand-waving is necessary anyway, one of the many instanton liquid models can be used to cure this. This is the subject of the following section.

\section{Landau background gauge}
\subsection{Computing the one-loop determinant} \label{oneloop}
In the background gauge, the LCO formalism must be slightly modified to remain renormalizable \cite{Vercauteren:2007gx}: the operator coupled to the source will now contain the quantum fluctations of the gauge field only: $\mathcal A_\mu^2 = (A_\mu-\hat A_\mu)^2$ with $\hat A_\mu^a$ the background field. This means that, at the classical level, when $\mathcal A_\mu^a$ is zero, the LCO formalism does not influence the equations. This means that the instanton will not be modified as in the non-background gauge in the previous section or as in electroweak theory ---in our case a vacuum expectation value for $\sigma$ will only give a mass to the quantum fluctuations, not to the classical part of $A_\mu$.

The computation of the one-loop quantum corrections to the action of massive fields in an instanton background is a non-trivial feat. Recently, Dunne \emph{et al.} have developed a strategy leading to an exact albeit numerical result \cite{dunneprl, dunnelang}. We give a short overview of the necessary steps as applied to spin and isospin 1 fields. More details can be found in \cite{dunnelang}.

We expand around a constant value for $\sigma$ and around a one-instanton configuration for $A_\mu^a$. The quantum fluctuation in $\sigma$ can be immediately integrated out, and we find that up to one-loop order:
\begin{multline}
V_\text{eff} = \frac{8\pi^2}{g^2} + V \frac{\sigma^2}{2 g^2\zeta} - \log\det(-\mathcal{D}^2) \\
+ \frac12 \log\det\left(- g_{\mu\nu} \mathcal{D}^2_{ab} + \left(1-\frac1\xi\right) (\mathcal{D}_\mu\mathcal{D}_\nu)^{ab} + 2g\epsilon^{abc} F_{\mu\nu}^c + \frac\sigma{g\zeta} g_{\mu\nu}\delta^{ab}\right)
\end{multline}
where all covariant derivatives contain only the instanton background, where the limit $\xi\rightarrow0$ for the Landau gauge is implied, and with $V$ the volume of space-time.

The $\log\det$ of the gluon propagator can be simplified as in 't~Hooft's original paper \cite{thooft}\footnote{'t~Hooft does not mention spin elimination for gluons, only for fermions, but the procedure is essentially the same.}. The longitudinal gluon turns out to remain massless, while the three other polarizations acquire a mass, as one would expect. The zero-modes must be treated separately. Due to the classical action being the unmodified Yang--Mills action, one would naively expect these modes to remain zero-modes. However, going through the computations uncovers that they get a mass $\sigma/g\zeta$, due to the perturbative approximation. Properly including all the interactions between $\sigma$ and the gluon field to all orders will make the zero-modes massless again, and we will treat them as such. Using the action without the Hubbard--Stratonovich transformation and with a source directly coupled to $\mathcal A_\mu^2$ shows that this is indeed the right course.

The remaining task is now to compute the functional determinant of the propagator of a massive particle in the presence of an instanton. Herefore the work by Dunne and collaborators can be followed. As a first step, the operators under consideration are separated into a radial part and an angular and isospin part. The angular and isospin quantum numbers couple according to the usual spin-orbit coupling mechanism, and we are left with one-dimensional operators within the determinants. An old trick relating the functional determinant of an operator $\hat{\mathcal O}$ to the asymptotic value of a function obeying $\hat{\mathcal O} f = 0$ can then be used, where the function $f$ will, in our case, be solved for numerically. In \cite{dunneprl,dunnelang} there is explained how to find a differential equation for the logarithm of the determinant itself, which is numerically more stable, and also how the convergence of the integration can be increased.

Now it remains to regularize the sum over the angular quantum numbers. This sum is, of course, divergent. A reorganization of the double sum makes it less divergent than the original one, but it does, of course, not yet completely solve the problem. In order to find a finite result, the theory has to be renormalized. Therefore we introduce a Pauli--Villars regulator. If we take a certain cut-off $l=L$ in our sum, we can separate it into two parts: one with $l\leq L$, where the Pauli--Villars regulator can be taken to infinity and which can be computed numerically to give a finite result, and one part with $l>L$, which has to be computed analytically and which must be used to subtract the large-$L$ divergence from the numerically determined sum.

\begin{figure}\begin{center}
\includegraphics[width=.5\textwidth]{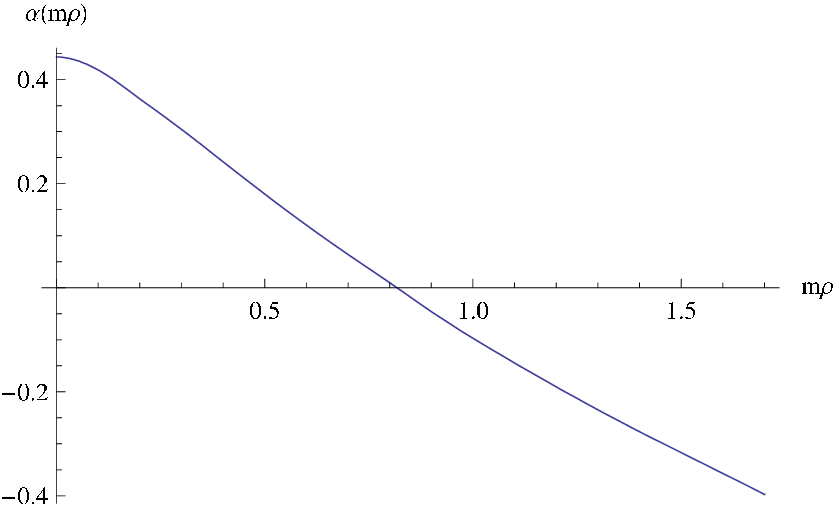}
\caption{The function $\alpha(m\rho)$ found from the computations in equation (3.2).}
\label{alpha}
\end{center} \end{figure}

This analytic computation can be done in a WKB expansion. In the limit of high $L$, only the first two orders in the WKB expansion contribute. This computation has been done by Dunne \emph{et al.} for particles in the fundamental representation of the gauge group, and the procedure can be straightforwardly applied to adjoint particles. Finally we find in dimensional regularization:
\begin{multline} \label{grotevergelijking}
\log\det\left(\frac{-\mathcal D^2 + \frac\sigma{g\zeta}}{-\partial^2 + \frac\sigma{g\zeta}}\right) = \frac13 \left(\frac2\epsilon+\ln\rho^2\bar\mu^2\right) +  \lim_{L\rightarrow\infty} \Bigg( \sum_{l=0,\frac12,1\ldots}^{L} \Gamma^S_l(\rho^2\sigma/g\zeta) + 8L^2 + 20L \\
- \ln L\left(\frac23+\frac{2\rho^2\sigma}{g\zeta}\right) + \frac{83}9 - \frac43\ln2 + \frac{\rho^2\sigma}{g\zeta} \left(2-4\ln2+\ln\frac{\rho^2\sigma}{g\zeta}\right) \Bigg)
\end{multline}
where $\Gamma_l(\rho^2\sigma/g\zeta)$ is the result from the numerical computation with quantum number $l$. Practically, taking $L\approx50$ gives acceptable results. The function defined by the limit of the expression between brackets is plotted in Figure \ref{alpha}.

Putting everything together and working in the dilute gas approximation, which sums all contributions from all numbers of instantons into an exponential, we find that the integration over the instanton size $\rho$ is divergent, as in the massless case. One might naively expect a gluon mass to cure this divergence, as happens in electroweak theory, but here the mass only enters in the quantum correction and does not operate at the classical level.

In order to extract meaningful results from the computation, the $\rho$-integral has to be given a finite value in some way. The easiest way out is to add an infrared cut-off $\rho_c$ as the upper bound of the integral, but this violates the scaling Ward identities \cite{Hutter:1995sc}. Several improvements have been suggested in the literature, usually involving interactions between the instantons. For our purpose, however, it suffices to take a phenomenological approach: we suppose the infrared divergence is somehow cured, and we work in an instanton liquid with certain values for the density $n$ and average radius $\rho$. This leads to the effective action
\begin{multline}
\frac1V V_\text{eff}(m^2,n,\rho) = \frac{27}{26} \frac{m^4}{2g^2} + \frac94 \frac{m^4}{(4\pi)^2} \left(-\frac56-\frac{161}{78}+\ln\frac{m^2}{\bar\mu^2}\right) \\
- n \exp\left(-\frac{8\pi^2}{g^2} + \frac{11}3 \ln\bar\mu^2\rho^2 - \frac32\alpha(m\rho) + \frac12\alpha(0)\right)
\end{multline}
where $m$ is the effective gluon mass defined as $m^2 = \frac\sigma{g\zeta} = \frac{N_c}{N_c^2-1}\frac{13}{9} g\sigma$, $\alpha(m\rho)$ is the function computed numerically and shown in Figure \ref{alpha}, and $\alpha(0) = - 8\zeta'(-1) - 10/9 + 1/3 \ln2$. Phenomenological values for $n$ and $\rho$ found on the lattice are \cite{Schafer:1996wv}
\begin{equation}
n \approx \unit{1}{\femto\meter^{-4}} \approx (\unit{0.6}{\lms})^4 \;, \qquad \rho \approx \unit{\frac13}{\femto\meter} \approx (\unit{1.8}{\lms})^{-1} \;,
\end{equation}
where $\lms = \unit{330}{\mega\electronvolt}$ in SU(2).

Taking the scale $\bar\mu^2$ at the value of $m^2$ in the global minimum of the action, we find that the instantons are much suppressed by the relative smallness of the coupling $g^2$. The non-perturbative minimum is still at $m\approx\unit{2.05}{\lms}$, as in the case without instantons. Now, however, we cannot say that $\langle \frac12g^2A_\mu^2\rangle = -\frac{27}{26}m^2 = -\unit{4.36}{\lms^2}$ in SU(2), since the instanton contribution to the condensate has to be included. As in \cite{Boucaud:2002nc,Boucaud:2002wy}\footnote{Mark that the authors of \cite{Boucaud:2002nc,Boucaud:2002wy} used a different convention for the gauge fields, and their $A_\mu^2$ corresponds to $g^2A_\mu^2$ here.} each instanton gives a contribution of $12\pi^2\rho^2$, resulting in
\begin{equation} \label{instantontotaal}
\langle g^2A_\mu^2\rangle_\text{tot} = -(\unit{2.0}{\lms})^2 = -\unit{0.42}{\giga\electronvolt^2} \;.
\end{equation}
This value depends strongly on the instanton liquid parameters plugged into the model. It is negative but close to zero because the instanton and quantum contributions are similar in magnitude but opposite in sign, and the quantum corrections have slightly larger absolute value.

\section{Conclusions} \label{conclusions}
A first conclusion arrived at in this paper is that we have not water-thightly been able to solve the infrared problem plaguing instanton physics by adding an effective gluon mass coming from the dimension two condensate. As the gluon mass must be determined from its gap equations, this leaves open the possibility of it being zero, which gives instantons the possibility to cause the infrared divergence. The amount of hand-waving necessary to stabilize the vacuum is less than without the condensate (one only has to state that the mass will be sufficiently high and the divergence is swept under the rug), but the state of affairs is not yet very satisfying.

The second main conclusion of this chapter is that, when working in the Landau background gauge, the LCO formalism gives a separate contribution to $\langle A_\mu^2\rangle$, which lowers the contributions coming from the instantons themselves.

\end{document}